\documentclass[
aps,%
12pt,%
final,%
notitlepage,%
oneside,%
onecolumn,%
nobibnotes,%
nofootinbib,%
superscriptaddress,%
showpacs,%
centertags]%
{revtex4}
\def\bb {\begin {eqnarray}}
\def\ee {\end {eqnarray}}

\begin{document}
\selectlanguage{english}
\title
{Integrable string models with constant torsion in terms of chiral\\
invariants of SU(n), SO(n), SP(n) groups }

\author{\firstname{V.D}~\surname{Gershun}}
\affiliation{ITP, NSC Kharkov Institute of Physics and Technology,
Kharkov,Ukraine}

\begin{abstract}
 We used the invariant local chiral currents of
principal chiral models for SU(n), SO(n), SP(n) groups to
construct new integrable string equations of hydrodynamic type on
the Riemmann space of the chiral primitive invariant currents and
on the chiral non-primitive Casimir operators as Hamiltonians.
\end{abstract}

\maketitle
\section{Introduction}\label{int1}
 For quantum
description of string model we must to have classical solutions of
the string in the background fields. String theory in suitable
space-time backgrounds can be considered as principal chiral
model. The integrability of the classical principal chiral model
is manifested through an infinite set of conserved charges, which
can form non-abelian algebra. Any charge from the commuting subset
of charges and any Casimir operators of charge algebra can be
considered as Hamiltonian. The polynomials of local chiral
currents was considered by Goldshmidt and Witten \cite{ger:Gol}
(see also \cite{ger:Are}). The local conserved chiral charges in
principal chiral models was considered by Evans, Hassan, MacKay,
Mountain \cite{ger:Eva}. The integrable string models of hydrodynamic
type was considered by author \cite{ger:Ger1}, \cite{ger:Ger2}.
In section \ref{sec2} the author obtained the basis local invariant chiral
currents for $SU(n)$ which they form closed algebra under Poisson brackets of hydrodynamic type.
The author constructed new invariant symmetric tensors and obtained relations between
new invariant tensors and basis tensors. In section \ref{sec3} the author
obtained new integrable string equations for $SU(3)$ group for chiral currents
of $SU(3)$ group as local fields of Riemmann space with non-primitive chiral
Casimirs as Hamiltonians. The non-primitive invariant chiral charges for $SU(n)$ $n \ge 4$
are not Casimirs and they can not used as Hamiltonians to construct new integrable equations
for this groups. The author show that new integrable string equations do not arise for
$SO(2l+1)$, $SO(2l)$, $SP(2l)$ groups for $l \ge 2$.

A string model is described by the Lagrangian
\[L=\frac{1}{2}\int\limits_{0}^{2\pi}\eta^{\alpha\beta}g_{ab} (\phi
(t,x))\frac{\partial \phi ^{a}(t,x)} {\partial
x^{\alpha}}\frac{\partial \phi ^{b}(t,x)}{\partial x^{\beta}}dx\]
and by two first kind constraints: \[g_{ab}(\phi
(x))[\frac{\partial\phi^{a}(x)}{\partial t}
\frac{\partial\phi^{b}(x)}{\partial
t}+\frac{\partial\phi^{a}(x)}{\partial
x}\frac{\partial\phi^{b}(x)} {\partial x}]\approx
0,\]\[g_{ab}(\phi (x))\frac{\partial\phi^{a}(x)}{\partial t}
\frac{\partial\phi^{b}(x)}{\partial x}\approx 0.\] The target
space local coordinates
$\phi^{a}(x),\,\,a=1,...,n$ belong to certain given smooth \\
$n$-dimensional manifold $M^{n}$ with nondegenerate metric tensor
\[g_{ab}(\phi(x))=\eta_{\mu\nu}e_{a}^{\mu}(\phi(x))
e_{b}^{\nu}(\phi(x)),\] where $\mu ,\nu =1,...,n$ are indexes of
tangent space to manifold $M^{n}$ on some point $\phi^{a}(x)$. The
veilbein $e_{a}^{\mu}(\phi)$ and its inverse $e_{\mu}^{a}(\phi)$
satisfy to the conditions: \[e_{a}^{\mu}e_{\mu}^{b}=\delta
_{a}^{b},\,\,e_{a}^{\mu}e^{a\nu } =\eta ^{\mu\nu}. \] The
coordinates $x^{\alpha}, x^{0}=t,\,x^{1}=x$ belong to world sheet
with metric tensor $g_{\alpha\beta}$ in conformal gauge.The string
equations of motion have the form:
\[\eta^{\alpha\beta}[\partial _{\alpha\beta}\phi^{a}+
\Gamma ^{a}_{bc}(\phi)\partial_{\alpha }
\phi^{b}\partial_{\beta}\phi^{c}]=0, \,\, \partial _{\alpha
}=\frac{\partial }{\partial x^{\alpha}}, \,\, \alpha =0,1, \] where
\[\Gamma^{a}_{bc}(\phi)=\frac{1}{2}e^{a}_{\mu}[\frac{\partial
e^{\mu}_{b}}{\partial \phi^{c}}+ \frac{\partial
e^{\mu}_{c}}{\partial \phi^{b}}]\] is connection.

The Hamiltonian has
form: \[H=\frac{1}{2}\int\limits_{0}^{2\pi}[\eta^{\mu\nu}
J_{0\mu}J_{0\nu}+\eta_{\mu\nu} J^{\mu}_{1}J^{\nu}_{1}]dx,\] where
$J_{0\mu}(\phi)=e^{a}_{\mu}(\phi)p_{a}$,
$J^{\mu}_{1}(\phi)=e^{\mu}_{a}\frac{\partial }{\partial
x}\phi^{a}$ and
$p_{a}(t,x)=\eta_{\mu\nu}e^{\mu}_{a}e^{\nu}_{b}\frac{\partial}{\partial
t}\phi^{b}$ is canonical momentum.

Let
us introduce chiral currents:
\[U^{\mu}=\eta^{\mu\nu}J_{0\nu}+J^{\mu}_{1},\,\,V^{\mu}=\eta^{\mu\nu}J_{0\nu}-
J^{\mu}_{1}\] The commutation relations of chiral currents
$U^{\mu}$ are following:
\begin{equation}\label{eq1}\{U^{\mu}(x),U^{\nu}(y)\}=C^{\mu\nu}_{\lambda}\,
[\frac{3}{2}U^{\lambda}(x)-\frac{1}{2} V^{\lambda}(x)\,]\delta
(x-y) -\eta^{\mu\nu}\frac{\partial}{\partial x}\delta
(x-y),\end{equation} where
\[C^{\mu}_{\nu\lambda}(\phi)=e^{a}_{\nu}e^{b}_{\lambda}[\frac{\partial
e^{\mu}_{a}}{\partial \phi^{b}} -\frac{\partial
e^{\mu}_{b}}{\partial \phi^{a}}]\] is torsion. Equations of motion
in light-cone coordinates
\[x^{\pm}=\frac{1}{2}(t\pm x),\,\,\frac{\partial }{\partial x^{\pm }}=
\frac{\partial}{\partial t}\pm \frac{\partial}{\partial x}\] have
form: \[\partial _{-}U^{\mu}=C^{\mu}_{\nu\lambda}(\phi(x))U^{\nu}
V^{\lambda},\,\,\partial _{-}V^{\mu}=
C^{\mu}_{\nu\lambda}(\phi(x))V^{\nu}U^{\lambda}.\] In the case of
null torsion: \[C^{\mu}_{\nu\lambda}=0,\,\,e^{\mu}_{a}(\phi)=
\frac{\partial e^{\mu}}{\partial \phi^{a}}, \,\,\Gamma
^{a}_{bc}(\phi)=e^{a}_{\mu}\frac{\partial ^{2}e^{\mu}}{{\partial
\phi^{b}}{\partial \phi^{c}}},\,\,
R^{\mu}_{\nu\lambda\rho}(\phi)=0\] string model is integrable
one.The Hamiltonian equations of motion under Hamiltonian $H$ are
described two independent left and right movers: $U^{\mu}(t+x)$
and $V^{\mu}(t-x)$.
\section{ Integrable string models with constant torsion}\label{sec2}
 Let us come back to commutations relations of chiral currents.
Let torsion $C^{\mu}_{\nu\lambda}(\phi(x))\ne 0$ and
$C_{\mu\nu\lambda}=f_{\mu\nu\lambda}$ are structure constant of
simple Lie algebra. We will consider string model with constant
torsion in light-cone gauge in target space. This model coincides
to principal chiral model on compact simple Lie group. We can not
divide motion on right and left mover because of chiral currents
$\partial_{-}U^{\mu}=f^{\mu}_{\nu\lambda}U^{\nu}V^{\lambda}$,
$\partial_{-}V^{\mu}= f^{\mu}_{\nu\lambda}V^{\nu}U^{\lambda}$ are
not conserve. The correspondent charges are not Casimirs. Evans,
Hassan, MacKay, Mountain constructed local invariant chiral
currents as polynomials of initial chiral currents of $SU(n)$,
$SO(n)$, $SP(n)$ principal chiral models and they found such
combination of them, that corresponding charges are Casimir
operators of this dynamical systems. Their paper was based on the
paper of de Azcarraga, Macfarlane, MacKay, Perez Bueno \cite{ger:Az}
about
invariant tensors for simple Lie algebras. Let $t_{\mu}$ are $n
\otimes n$ traceless hermitian matrix representations of
generators Lie algebra:
\[[t_{\mu},t_{\nu}]=2if_{\mu\nu\lambda}t_{\lambda} , \,\, Tr
(t_{\mu}t_{\nu})=2\delta_{\mu\nu}.\] Here is additional relation
for $SU(n)$ algebra:
\[\{t_{\mu},t_{\nu}\}=\frac{4}{n}\delta_{\mu\nu}+
2d_{\mu\nu\lambda} t_{\lambda}, \,\,\mu=1,...,n^{2}-1 .\] De
Azcarraga et. al. gave some examples of invariant tensors of
simple Lie algebras and they gave general method to calculate
them. Invariant tensors may to construct as invariant symmetric
polynomials on $SU(n)$:
\[d_{M}=d_{(\mu_{1}...\mu_{M})}= \frac
{1}{M!}STr(t_{\mu_{1}}...t_{\mu_{M}}),\] where $STr $ means of
completely symmetrized product of matrices and
$d_{(\mu_{1}...\mu_{M})}$ is totally symmetric tensor and
$M=2,3,...,\infty $. Another family of invariant symmetric
tensors, such named $d$-family , based on the product of the
symmetric structure constant $d_{\mu\nu\lambda}$ of $SU(n)$
algebra: \[C_{M}=d_{(\mu_{1}...\mu_{M})}=
d^{k_{1}}_{({\mu_{1}\mu_{2}}}d^{k_{1}k_{2}}_{\mu_{3}}...d^{k_{M-2}
k_{M-3}}_{\mu_{M-2}}d^{k_{M-3}}_{\mu_{M-1}\mu_{M})},\] where
$C_{3}=d_{\mu\nu\lambda}$ and $M=4,5,...,\infty.$

Here are $n-1$ primitive invariant tensors on $SU(n)$. The
invariant tensors for $M \ge n$ are functions of primitive
tensors. Evans et.al. introduced local chiral currents based on
the invariant symmetric polynomials on simple Lie groups:
\[J_{M}(U)=d_{\mu_{1}...\mu_{M}}U^{\mu_{1}}...U^{\mu_{M}},\] where
$U=t_{\mu}U^{\mu}$ and $\mu =1,...,n^{2}-1$ . It is possible to
compose the invariant symmetric polynomials $J_{M}(U)$ to basis
invariant tensors $C_{M}(U)$:
\begin{equation}\label{eq2}C_{2}(U)=\eta_{\mu\nu}U^{\mu}U^{\nu},\,C_{3}(U)=d_{\mu\nu\lambda}
U^{\mu}U^{\nu}U^{\lambda},\,\, C_{M}(U)=
d^{k_{1}}_{\mu_{1}\mu_{2}}d^{k_{1}k_{2}}_{\mu_{3}}...
 d^{k_{M-3}}_{\mu_{M-1}\mu_{M}}
U_{\mu_{1}}U_{\mu_{2}}...U_{\mu_{M}},\end{equation} where
$M=4,5,...,\infty .$
 The author obtained following expression for local invariant chiral currents $J_{M}(U)$:
\begin{equation}J_{2}=2C_{2}, \,\, J_{3}=2C_{3},\,\,
J_{4}=2C_{4}+\frac{4}{n}C_{2}^{2}, \,\,
J_{5}=2C_{5}+\frac{8}{n}C_{2}C_{3},\nonumber\end{equation}
\begin{equation}\label{eq3}J_{6}=2C_{6}+\frac{4}{n}C_{3}^{2}+\frac{8}{n}
C_{2}C_{4}+ \frac{8}{n^{2}}C_{2}^{3},\end{equation}
\begin{equation}J_{7}=2C_{7}+\frac{8}{n}C_{3}C_{4}+\frac{8}{n}C_{2}C_{5}+
\frac{24}{n^{2}}C_{2}^{2}C_{3},\nonumber\end{equation}
\[J_{8}=2C_{8}+\frac{4}{n}C_{4}^{2}+\frac{8}{n}C_{3}C_{5}+
\frac{8}{n}C_{2}C_{6}+\frac{24}{n^{2}}C_{2}C_{3}^{2}+\frac{24}{n^{2}}C_{2}^{2}C_{4}+\frac{16}
{n^{3}}C_{2}^{4},\]
\[J_{9}=2C_{9}+\frac{8}{n}C_{4}C_{5}+\frac{8}{n}C_{3}C_{6}+\frac{8}{n}C_{2}C_{7}
+\frac{8}{n^{2}}C_{3}^{3}+\frac{48}{n^{2}}C_{3}C_{4}+
\frac{24}{n^{2}}C_{2}^{2}C_{5}+\frac{64}{n^{3}}C_{2}^{3}C_{3}.\]

 The commutation relations of invariant chiral currents $J_{M}(U(x))$ show, that
these currents are not densities of dynamical Casimir operators.
 We considered
the basis family of invariant chiral currents $C_{M}(U)$ and we
proved that invariant chiral currents $C_{M}(U)$ forms closed
algebra under canonical PB and corresponding charges are dynamical
Casimir operators. The commutation relations of invariant chiral
currents $C_{M}(U(x))$ and $C_{N}(U(y))$ for $M,N = 2,3,4$ and for
$M=2, N=2,3,...,\infty$ are the following:
\[\{C_{M}(x),C_{N}(y)\}=-MN C_{M+N-2}(x) \frac{\partial }
{\partial x}\delta (x-y)-\frac{MN(N-1)}{M+N-2}\frac{\partial
C_{M+N-2}(x)} {\partial x}\delta(x-y).\]
The commutations
relations for $M \ge 5, N \ge 3$ are following (we show one
formula for $C_{8}$ on the right side as example only):
\[\{C_{5}(x),C_{3}(y)\}=-[12C_{6}(x)+3C_{6,1}(x)]\frac{\partial }{\partial x}
\delta (x-y) - \frac{1}{3}\frac{\partial}{\partial
x}[12C_{6}(x)+3C_{6,1}(x)]\delta (x-y),\]
\[\{C_{5}(x),C_{4}(y)\}=-[16C_{7}(x)+4C_{7,1}(x)]\frac{\partial }{\partial x}
\delta (x-y) - \frac{3}{7}\frac{\partial}{\partial
x}[16C_{7}(x)+4C_{7,1}(x)]\delta (x-y),\]
\[\{C_{6}(x),C_{3}(y)\}=-[12C_{7}(x)+6C_{7,1}(x)]\frac{\partial }{\partial x}
\delta (x-y) - \frac{2}{7}\frac{\partial}{\partial
x}[12C_{7}(x)+6C_{7,1}(x)]\delta (x-y),\]
\[\{C_{5}(x),C_{5}(y)\}=-[16C_{8}(x)+8C_{8,1}(x)+C_{8,2}(x)]\frac{\partial }
{\partial x}\delta (x-y)-\]\[\frac{1}{2}\frac{\partial}{\partial x}[16C_{8}(x)+8C_{8,1}(x)+C_{8,2}(x)]
\delta (x-y),\]
\begin{equation}\label{eq4}\{C_{6}(x),C_{4}(y)\}=-[16C_{8}(x)+8C_{8,3}(x)]\frac{\partial }
{\partial x}\delta (x-y)-\end{equation}\[\frac{3}{8}\frac{\partial}{\partial
x}[16C_{8}(x)+8C_{8,3}(x)] \delta (x-y),\]
\[\{C_{7}(x),C_{3}(y)\}=-[12C_{8}(x)+6C_{8,1}(x)+3C_{8,3}]\frac{\partial }
{\partial x}\delta (x-y)-\]\[\frac{1}{4}\frac{\partial}{\partial x}[12C_{8}(x)+6C_{8,1}(x)+3C_{8,3}(x)]
\delta (x-y),\]
\[\{C_{8}(x),C_{3}(y)\}=-[12C_{9}(x)+6C_{9,1}(x)+6C_{9,2}(x)]\frac{\partial }
{\partial x}\delta (x-y)
-\]\[\frac{2}{9}\frac{\partial}{\partial x}[12C_{9}(x)+6C_{9,1}(x)+6C_{9,2}(x)]
\delta (x-y),\]
\[\{C_{7}(x),C_{4}(y)\}=-[16C_{9}(x)+8C_{9,2}(x)+4C_{9,3}(x)]\frac{\partial }
{\partial x}\delta (x-y)
-\]\[\frac{1}{3}\frac{\partial}{\partial x}[16C_{9}(x)+8C_{9,2}(x)+4C_{9,3}(x)]
\delta (x-y),\]
\[\{C_{6}(x),C_{5}(y)\}=-[16C_{9}(x)+4C_{9,1}(x)+8C_{9,2}(x)+2C_{9,4}(x)]
\frac{\partial }{\partial x}\delta (x-y)
-\]\[\frac{4}{9}\frac{\partial}{\partial x}[16C_{9}(x)+4C_{9,1}(x)+8C_{9,2}(x)+
2C_{9,4}(x)]\delta (x-y),\]
\[\{C_{9}(x),C_{3}(y)\}=-[12C_{10}(x)+6C_{10,1}(x)+6C_{10,2}(x)+3C_{10,3}(x)]
\frac{\partial }{\partial x}\delta (x-y)
-\]\[\frac{1}{5}\frac{\partial}{\partial x}[12C_{10}(x)+6C_{10,1}(x)+6C_{10,2}(x)+
3C_{10,3}(x)]\delta (x-y),\]
\[\{C_{8}(x),C_{4}(y)\}=-[16C_{10}(x)+8C_{10,2}(x)+8C_{10,4}(x)]
\frac{\partial }{\partial x}\delta (x-y)
-\]\[\frac{3}{10}\frac{\partial}{\partial x}[16C_{10}(x)+8C_{10,2}(x)+
8C_{10,4}(x)]\delta (x-y),\]
\[\{C_{7}(x),C_{5}(y)\}=-[16C_{10}(x)+8C_{10,3}(x)+4C_{10,1}(x)+4C_{10,4}(x)+
2C_{10,5}(x)+C_{10,6}(x)] \frac{\partial }{\partial x}\delta (x-y)
-\]\[\frac{2}{5}\frac{\partial}{\partial x}[16C_{10}(x)+8C_{10,3}(x)+4C_{10,1}(x)+
4C_{10,4}(x)+2C_{10,5}(x)+C_{10,6}(x)]\delta (x-y),\]
\[\{C_{6}(x),C_{6}(y)\}=-[16C_{10}(x)+16C_{10,2}(x)+4C_{10,7}(x)]
\frac{\partial }{\partial x}\delta (x-y)
-\]\[\frac{1}{2}\frac{\partial}{\partial x}[16C_{10}(x)+16C_{10,2}(x)+
4C_{10,7}(x)]\delta (x-y).\]
%\begin{equation}\label{eq4}\{C_{6}(x),C_{4}(y)\}=-[16C_{8}(x)+8C_{8,3}(x)]\frac{\partial }
%{\partial x}\delta (x-y)-\frac{3}{8}\frac{\partial}{\partial
%x}[16C_{8}(x)+8C_{8,3}(x)] \delta (x-y),\end{equation}

 The new dependent invariant chiral currents $C_{6,1}$, $C_{7,1}$, $C_{8,1}- C_{8,3}$,
$C_{9,1}-C_{9,4}$, $C_{10,1}-C_{10,7}$ have the similar form (we show
formulas for $C_{8}$ on the right side as example only):
\[C_{8,1}=[d^{k}_{\mu\nu}d^{kl}_{\lambda}d^{ln}_{\rho}][d^{m}_{\sigma\varphi}]
[d^{p}_{\tau\theta}]d^{nmp}(U^{8})^{\mu\nu\lambda\rho\sigma\varphi\tau\theta},\]
\[C_{8,2}=[d^{k}_{\mu\nu}][d^{l}_{\lambda\rho}][d^{n}_{\sigma\varphi}]
[d^{m}_{\tau\theta}]d^{klp}d^{nmp}(U^{8})^{\mu\nu\lambda\rho\sigma\varphi\tau\theta}.\]
\[C_{8,3}=[d^{k}_{\mu\nu}d^{kl}_{\lambda}][d^{n}_{\rho\sigma}d^{nm}_{\varphi}]
[d^{p}_{\tau\theta}]d^{lmp}(U^{8}){\mu\nu\lambda\rho\sigma\varphi\tau\theta},\]

 The following picture show the graphic image of invariant chiral currents $C_{M}(U)$ and
the difference between $C_{8}$ and $C_{81}-C_{83}$.
\begin{figure}[h]
\setcaptionmargin{5mm} \onelinecaptionstrue
\includegraphics[scale=1]{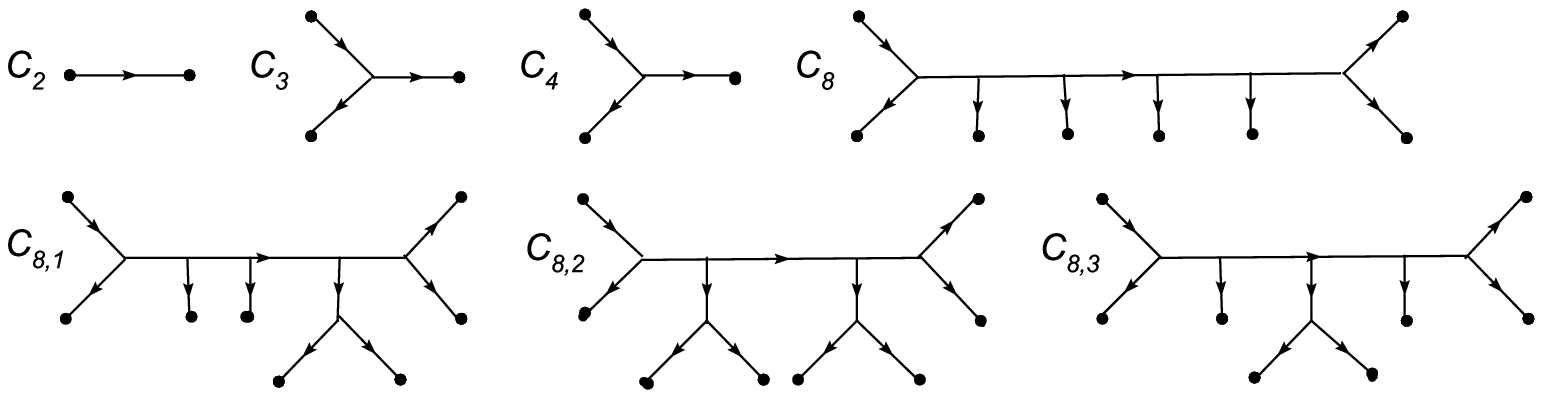}
\end{figure}

 Let us note that these PB's (\ref{eq4}) are PB's of hydrodynamic type. The
ultra local term with antisymmetric structure constant
$f_{\mu\nu\lambda}$ in commutation relation of chiral currents
$U^{\mu}$ (\ref{eq1})  does not contribution to commutation
relations of invariant chiral currents because of totally
symmetric invariant tensors $d_{(\mu_{1}...\mu_{M})}$. Therefore
chiral currents $C_{M}(U(x))$ form closed algebra under canonical
PB.

 The
new dependent invariant chiral currents and the new dependent
totally symmetric invariant tensors for $SU(N)$ group can be
obtained under different order of calculation of trace of the
product of the generators of $SU(n)$ algebra. Let us mark the
matrix product of two generators $t_{\mu}$, $t_{\nu}$ in round
brackets:
\[(t_{\mu}t_{\nu})=\frac{2}{n}\delta_{\mu\nu}+(d^{k}_{\mu\nu}+if^{k}_{\mu\nu})
t_{k}.\] The expression of invariant chiral currents $J_{M}(U)$
depends of the order of the matrix product of two generators in
general list of generators. For example:
\[ J_{8}=Tr[t(\underline{tt})tt(\underline{tt})t]=2C_{8}+\frac{4}{n}C_{4}^{2}+\frac{8}{n}C_{3}C_{5}+
\frac{8}{n}C_{2}C_{6}+
\frac{24}{n^{2}}C_{2}C_{3}^{2}+\frac{24}{n^{2}}C_{2}^{2}C_{4}+\frac{16}
{n^{3}}C_{2}^{4},\]
\[ J_{8}=Tr[(\underline{tt})(\underline{tt})t(\underline{tt})t]=2C_{8,1}+\frac{4}{n}C_{4}^{2}
+\frac{4}{n}C_{3}C_{5}+\frac{24}{n^{2}}C_{2}
C_{3}^{2}+\frac{12}{n}C_{2}C_{6}+
\frac{24}{n^{2}}C_{2}^{2}C_{4}+\frac{16}{n^{3}}C_{2}^{4},\]
\[ J_{8}=Tr[(\underline{tt})(\underline{tt})(\underline{tt})(\underline{tt})]=
2C_{8,2}+\frac{4}{n}C_{4}^{2}+\frac{16}{n}C_{2}
C_{6,1}+\frac{32}{n^{2}}C_{2}^{2}C_{4}+\frac{16}{n^{3}}C_{2}^{4},\]
\[ J_{8}=Tr[t(\underline{tt})(\underline{tt})(\underline{tt})t]=2C_{8,3}+
\frac{12}{n}C_{2}C_{6}+\frac{8}{n}C_{3}C_{5}+\frac{24}{n^{2}}C_{2}^{2}C_{4}+
\frac{24}{n^{2}}C_{2}C_{3}^{2}+ \frac{16}{n}C_{2}^{4},\]
\[J_{9}=Tr[t(tt)ttt(tt)t]=2C_{9}+\frac{8}{n}C_{4}C_{5}+\frac{8}{n}C_{3}C_{6}+
\frac{8}{n}C_{2}C_{7}+\frac{8}{n^{2}}C^{3}_{3}+
\frac{48}{n^{2}}C_{2}C_{3}C_{4}+\frac{24}{n^{2}}C^{2}_{2}C_{5}+\frac{64}{n^{3}}
C^{3}_{2}C_{3},\]
\[J_{9}=Tr[t(tt)tt(tt)(tt)]=\]
$$
\left\{\begin{array}{lll}2C_{9,1}+\frac{4}{n}C_{4}C_{5}+\frac{4}{n}C_{2}C_{7}+
\frac{4}{n}C_{2}C_{7,1}+ \frac{8}{n}C_{3}C_{6,1}+
\frac{32}{n^{2}}C_{2}C_{3}C_{4}+\frac{32}{n^{2}}C^{2}_{2}C_{5}+\frac{64}{n^{3}}
C^{3}_{2}C_{3} \bigskip\\
2C_{9,4}+\frac{4}{n}C_{2}C_{7}+ \frac{4}{n}C_{2}C_{7,1}+
\frac{12}{n}C_{3}C_{6,1}+
\frac{32}{n^{2}}C_{2}C_{3}C_{4}+\frac{32}{n^{2}}C^{2}_{2}C_{5}+\frac{64}{n^{3}}
C^{3}_{2}C_{3},\\ \end{array}\right.
$$
\[J_{9}=Tr[t(tt)t(tt)t(tt)]=\]
$$\left\{\begin{array}{lll}2C_{9,2}+\frac{4}{n}C_{4}C_{5}+\frac{8}{n}C_{3}C_{6}+
\frac{8}{n}C_{2}C_{7}+\frac{4}{n}C_{2}C_{7,1}
+\frac{8}{n^{2}}C^{3}_{3}+\frac{40}{n^{2}}C_{2}C_{3}C_{4}+\frac{32}{n^{2}}
C^{2}_{2}C_{5}+\frac{64}{n^{3}}C^{3}_{2}C_{3},\bigskip \\
2C_{9,3}+\frac{8}{n}C_{2}C_{7}+\frac{4}{n}C_{2}C_{7,1}+\frac{12}{n}C_{3}C_{6}+
\frac{8}{n}C^{3}_{3}+
\frac{40}{n^{2}}C_{2}C_{3}C_{4}+\frac{32}{n^{2}}C^{2}_{2}C_{5}+\frac{64}{n^{3}}
C^{3}_{2}C_{3},\\ \end{array}\right. $$

 where $t=t_{\mu}U^{\mu}$ and two variants of two last expressions for $J_{9}(U)$
obtained from two variants of expression for $J_{6}(U)$ during calculation $J_{9}(U)$.
 Because result of calculation does not depend
of order of calculation, we can obtain relations between new
invariant chiral currents and basis invariant currents $C_{M}(U)$:
\[ C_{6,1}=C_{6}+\frac{2}{n}C^{2}_{3}-\frac{2}{n}C_{2}C_{4},\]
\[ C_{7,1}=C_{7}+\frac{4}{n}C_{3}C_{4}-\frac{4}{n}C_{2}C_{5},\]
\begin{equation}\label{eq5} C_{8,1}=C_{8}+\frac{2}{n}C_{3}C_{5}-\frac{2}{n}C_{2}C_{6},\end{equation}
\[ C_{8,2}=C_{8}+\frac{4}{n}C_{3}C_{5}-\frac{4}{n}C_{2}C_{6}-\frac{4}{n^{2}}C_{2}C^{2}_{3}+\frac{4}{n^{2}}C^{2}_{2}C_{4},\]
\[ C_{8,3}=C_{8}+\frac{2}{n}C^{2}_{4}-\frac{2}{n}C_{2}C_{6},\]
\[ C_{9,1}=C_{9}+\frac{2}{n}C_{4}C_{5}-\frac{4}{n^{2}}C^{3}_{3}+\frac{8}{n^{2}}C_{2}C_{3}C_{4}+\frac{4}{n^{2}}C^{2}_{2}C_{5},\]
\[ C_{9,2}=C_{9}+\frac{2}{n}C_{4}C_{5}-\frac{2}{n}C_{2}C_{7}-\frac{4}{n^{2}}C_{2}C_{3}C_{4}+\frac{4}{n^{2}}C^{2}_{2}C_{5},\]
\[ C_{9,3}=C_{9}+\frac{4}{n}C_{4}C_{5}-\frac{2}{n}C_{2}C_{7}- \frac{2}{n}C_{3}C_{6}-\frac{4}{n^{2}}C_{2}C_{3}C_{4}+
\frac{4}{n^{2}}C^{2}_{2}C_{5},\]
\[ C_{9,4}=C_{9}+\frac{4}{n}C_{4}C_{5}-\frac{2}{n}C_{3}C_{6}-\frac{8}{n^{2}}C^{3}_{3}+\frac{12}{n^{2}}C_{2}C_{3}C_{4}+
\frac{4}{n^{2}}C^{2}_{2}C_{5}.\]
 Hence we can obtain the new relations for symmetric tensors:
\[d^{k}_{(\mu\nu}d^{l}_{\lambda\rho}d^{n}_{\sigma\varphi}d^{nm}_{\tau}d^{mp}_{\theta)}d^{klp}=
d^{k}_{(\mu\nu}d^{kl}_{\lambda}d^{ln}_{\rho}d^{nm}_{\sigma}d^{mp}_{\varphi}d^{p}_{\tau\theta
)}+
\frac{4}{n}d_{(\mu\nu\lambda}d^{k}_{\rho\sigma}d^{kl}_{\varphi}d^{l}_{\tau\theta)}-
\frac{2}{n}\delta_{(\mu\nu}d^{k}_{\lambda\rho}d^{kl}_{\sigma}d^{ln}_{\varphi}d^{n}_{\tau\theta
)},\]
\[d^{k}_{(\mu\nu}d^{l}_{\lambda\rho}d^{n}_{\sigma\varphi}d^{m}_{\tau\theta)}d^{klp}d^{nmp}=
d^{k}_{(\mu\nu}d^{kl}_{\lambda}d^{ln}_{\rho}d^{nm}_{\sigma}d^{mp}_{\varphi}d^{p}_{\tau\theta)}+
\frac{4}{n}d_{(\mu\nu\lambda}d^{k}_{\rho\sigma}d^{kl}_{\varphi}d^{l}_{\tau\theta)}-\]
\[-\frac{4}{n}\delta_{(\mu\nu}d^{k}_{\lambda\rho}d^{kl}_{\sigma}d^{ln}_{\varphi}
d^{n}_{\tau\theta)}-\frac{4}{n^{2}}\delta_{(\mu\nu}d_{\lambda\rho\sigma}d_{\varphi\tau\theta)}+
\frac{4}{n^{2}}\delta_{(\mu\nu}\delta_{\lambda\rho}d^{k}_{\sigma\varphi}d^{k}_{\tau\theta)}.\]

 The family of invariant chiral
currents $C_{M}(U(x))$ satisfy to conservation equations
$\partial_{-} C_{M}(U(x))=0$.
\section{New integrable string equations}\label{sec3}
 Let us apply hydrodynamic approach to integrable string models
with constant torsion. In this case we must to consider the
conserved primitive chiral currents currents $C_{M}(U(x))$,
$(M=2,3,...,n-1)$  as local fields of the Riemmann manifold. The
non-primitive local charges of invariant chiral currents with $M
\ge n$ form the hierarchy of new Hamiltonians in bi-Hamiltonian
approach to integrable systems. The commutation relations of
invariant chiral currents are local PBs of hydrodynamic type.

 The invariant chiral currents $C_{M}$ with $M \ge 3$ for the
$SU(3)$ group can be obtained from following relation:
\[d_{kln}d_{kmp}+d_{klm}d_{knp}+d_{klp}d_{knm}=
\frac{1}{3}(\delta_{ln}\delta_{mp}+\delta_{lm}\delta_{np}+\delta_{lp}
\delta_{nm}).\]

 The corresponding invariant chiral currents for $SU(3)$ group have
form:
\[C_{2N}=\frac{1}{3^{N-1}}(\eta_{\mu\nu}U^{\mu}U^{\nu})^{N}=
\frac{1}{3^{N-1}}(C_{2})^{N},\] \[
C_{2N+1}=\frac{1}{3^{N-1}}(\eta_{\mu\nu}U^{\mu}U^{\nu})^{N-1}d_{kln}U^{k}U^{l}U^{n}=
\frac{1}{3^{N-1}}(C_{2})^{N-1}C_{3}.\] The invariant chiral
currents $C_{2}, C_{3}$ are local coordinates of the Riemmann
manifold $M^{2}$. The local charges $C_{2N}$, $N \ge 2$ form
hierarchy of Hamiltonians. The new nonlinear equations of motion
for chiral currents are the following:
\begin{equation}\label{eq6}\frac{\partial f(x,t_{N})}{\partial
t_{N}}= -{f}^{N-1}(x)\frac{\partial f(x)}{\partial x},\end{equation}
\[\frac{\partial g(x,t_{N})}{\partial t_{N}}=
-\frac{3}{2N-1}g(x)f^{N-2}(x)\frac{\partial f(x)}{\partial
x} -\frac{1}{2N-1}f^{N-1}(x)\frac{\partial g(x)}{\partial
x}.\] Here $f(x)=C_{2}(U(x))$, $g(x)=C_{3}(U(x))$ and Hamiltonian
has following form:
\[H_{N}(t_{N})= \frac{1}{2N(2N-1)}\int\limits_{0}^{2\pi}f^{N}(x,t_{N})dx,\,\, N=2,...,\infty .\]
First equation for function $f(x,t_{n})$ is generalized inviscid
Burgers' equation. It has following solution:
\begin{equation}\label{eq7}f(x,t_{N})=h[x-t_{N}f^{N-1}(x,t_{N})]\end{equation}
and $h(x)$ is arbitrary function of $x$.

 The construction of integrable equations with $SU(n)$ symmetries for $n \ge 4$
has difficulties of reduction non-primitive invariant currents to
primitive currents. Following Balog, Feher, O'Raifeartaigh,
Forgacs, Wipf we considered the generating function
\[A(x,\lambda)=det(1-\lambda U(x))=\exp {Tr \ln (1-\lambda U^{\mu}(x)t_{\mu})}.\]
We obtained following expressions for non-primitive invariant
currents $C_{N}$:
\[SU(4): C_{5}\to \frac{2}{3}C_{2}C_{3},\,C_{6}\to
\frac{1}{6}C_{3}^{2}+\frac{1}{2}C_{2}C_{4}, \,\,C_{7}\to
\frac{1}{3}C_{2}^{2}C_{3}+\frac{1}{6}C_{3}C_{4},\]
\[C_{8}\to
\frac{7}{36}C_{2}C_{3}^{2}+\frac{1}{4}C_{2}^{2}C_{4}, \,\,C_{9}\to
\frac{1}{6}C_{2}^{3}C_{3}+\frac{1}{36}C_{3}^{3}+\frac{1}{6}C_{2}C_{3}C_{4}.\]
\[SU(5): C_{6}\to
-\frac{3}{50}C_{2}^{3}+\frac{4}{15}C_{3}^{2}+\frac{7}{10}C_{2}C_{4}, \,\,
C_{7}\to
-\frac{3}{50}C_{2}^{2}C_{3}+\frac{11}{30}C_{3}C_{4}+\frac{3}{5}C_{2}C_{5},\]
\[C_{8} \to
-\frac{9}{250}C_{2}^{4}\frac{4}{25}C_{2}C_{3}^{2}+\frac{9}{25}C_{2}^{2}C_{4}+
\frac{1}{10}C_{4}^{2}+\frac{4}{15}C_{3}C_{5},\]
\[C_{9}\to
-\frac{13}{250}C_{2}^{3}C_{3}+\frac{16}{225}C_{3}^{3}+\frac{61}{150}C_{2}C_{3}C_{4}
+\frac{3}{10}C_{2}^{2}C_{5}+\frac{1}{10}C_{4}C_{5}.\]

 However, the non-primitive charges are not commuting. They are
 not Casimirs and we can not consider them as Hamiltonians.
The similar method of construction chiral currents for
$SO(2l+1)=B_{l}$, $SP(2l)=C_{l}$ groups was used by Evans et.al.
on the base of symmetric invariant tensors of de Azcarraga et.al..
In the defining representation these groups generators
corresponding algebras $t_{\mu}$ satisfy rules:
\[ [t_{\mu},t_{\nu}]=2if_{\mu\nu}^{\lambda}t_{\lambda},\,
Tr(t_{\mu}t_{\nu})=2\delta_{\mu\nu},\,t_{\mu}\eta=-\eta
t_{\mu}^{t},\] where $\eta$ is euclidean or symplectic structure.

The symmetric tensor ctructure constants for these groups they
introduced through completely symmetrized product of three
generators of corresponding algebras:
\begin{equation}t_{(\mu}t_{\nu}t_{\lambda )}=
v^{\rho}_{\mu\nu\lambda}t_{\rho},\end{equation} where
$v_{\mu\nu\lambda\rho}$ is totally symmetric tensor.
 The basic invariant
symmetric tensors have the form :
\[V^{(2)}_{\mu\nu}=\delta_{\mu\nu}, \,\,
V^{(2N)}_{(\mu_{1}\mu_{2}...\mu_{2N-1}\mu_{2N})}=
v^{\nu_{1}}_{(\mu_{1}\mu_{2}\mu_{3}}v^{\nu_{1}\nu_{2}}_{\mu_{4}\mu_{5}}...
v^{\nu_{2N-3}}_{\mu_{2N-2}\mu_{2N-1}\mu_{2N})}, \,\, N=2,...,\infty
.\] The invariant chiral currents $J^{(2N)}$ coincide to the basis
invariant chiral currents $V^{(2N)}$ :
\[J^{(2N)}=2V^{(2N)}_{\mu_{1}...\mu_{2N}}U^{\mu
_{1}}...U^{\mu_{2N}}.\] The commutation relations of invariant
chiral currents are PBs of hydrodynamic type:
\[
\{ V^{(M)}(x),V^{(N)}(y)\}=-MN V^{(M+N-2)}(x)\frac{\partial
}{\partial x}\delta(x-y)-\frac{MN(N-1)}{M+N-2}\frac{\partial
V^{(M+N-2)}(x)} {\partial x}\delta(x-y).\] The commuting charges
of these  invariant chiral currents are dynamical Casimir
operators on $SO(2l+1)$, $SP(2l)$. The metric tensor of Riemmann
space of invariant chiral currents is the following:
\[g_{MN}(V(x))=-MN(M+N-2) V^{(M+N-2)}(x).\]
 The commutation relations coincide to commutation relations, which was obtained
by Evans at.al. .

 We used relations for new symmetric invariant tensors $V^{(2N,1)}_{(\mu_{1}...\mu_{2N})}$ ,
which we obtained during calculation PB :
\[V^{(10,1)}_{(\mu_{1}...\mu_{10})}=v_{(\mu_{1}\mu_{2}\mu_{3}}^{k}v_{\mu_{4}\mu_{5}\mu_{6}}^{l}v_{\mu_{7}\mu_{8}\mu_{9}}^{n}
v_{\mu_{10})}^{kln}=
V^{(10)}_{(\mu_{1}...\mu_{10})},\]
\[V^{(12,1)}_{(\mu_{1}...\mu_{12})}= v_{(\mu_{1}\mu_{2}\mu_{3}}^{k}v_{\mu_{4}\mu_{5}\mu_{6}}^{l}v_{\mu_{7}\mu_{8}\mu_{9}}^{n}
v_{\mu_{10}\mu_{11}\mu_{12})}^{m}v^{klnm}=V^{(12)}_{(\mu_{1}...\mu_{12})},\]
\[ V^{(14,1)}_{(\mu_{1}...\mu_{14})}=v_{(\mu_{1}\mu_{2}\mu_{3}}^{k}v_{\mu_{4}\mu_{5}\mu_{6}}^{l}v_{\mu_{7}\mu_{8}\mu_{9}}^{n}
v_{\mu_{10}\mu_{11}\mu_{12}}^{m}v_{\mu_{13}}^{klp}v_{\mu_{14})}^{nmp}=
V^{(14)}_{(\mu_{1}...\mu_{14})}.\] The invariant chiral currents
$J^{(2N)}$ and $V^{(2N)}$ and the new dependent totally symmetric
invariant tensors for $SO(2l+1)$, $SP(2l)$ groups can be obtained
under different order of calculation of trace of the product of
the generators of corresponding algebras.
 Let us mark the matrix product of three generators $t_{\mu}$ in round brackets:
\[(t_{(\mu}t_{\nu}t_{\lambda)})=v_{\mu\nu\lambda\rho}t_{\rho}.\]
The different disposition of this triplet inside of $J^{2N}$
produce different expressions for $V^{2N}$:
\[J^{(10)}=Tr[((\underline{ttt})t(\underline{ttt})(\underline{ttt}))]=2V^{(10)}, \,\,
J^{(10)}=Tr[((\underline{ttt})(\underline{ttt})(\underline{ttt})t]=2V^{(10,1)},\]
\[J^{(12)}=Tr[t(\underline{ttt})t(\underline{ttt})t(\underline{ttt})]=2V^{(12)}, \,\,
J^{(12)}=Tr[((\underline{ttt})(\underline{ttt})(\underline{ttt})(\underline{ttt}))]=2V^{(12,1)},\]
\[J^{(14)}=Tr[((\underline{ttt})t(\underline{ttt})t(\underline{ttt})(\underline{ttt}))]=2V^{(14)}, \,\,
[J^{(14)}=Tr[((\underline{ttt})(\underline{ttt})(\underline{ttt})(\underline{ttt})tt)]=2V^{(14,1)}.\]

 New invariant chiral tensors do not led to new invariant chiral
currents. We obtained following expressions for non-primitive
invariant currents $V_{N}$ for $SO(2l+1)$, $SO(2l)$, $SP(2l)$
groups:
\[l=1: V_{2N}\to V_{2}^{N},\]
\[l=2: V_{6}\to -\frac{1}{2}V^{3}_{2}+\frac{3}{2}V_{2}V_{4},\,\,
V_{8}\to -\frac{1}{2}V^{4}_{2}+V_{2}^{2}V_{4}+\frac{1}{2}V_{4}^{2},\,\,
V_{10}\to -\frac{1}{4}V_{2}^{5}+\frac{5}{4}V_{2}V_{4}^{2},\]
\[l=3: V_{8}\to
\frac{1}{6}V_{2}^{4}-V_{2}^{2}V_{4}+\frac{1}{2}V_{4}^{2}+\frac{4}{3}V_{2}V_{6},\,\,
V_{10}\to \frac{1}{6}V_{2}^{5}-\frac{5}{2}(V_{2}^{3}V_{4}-V_{2}^{2}V_{6}-V_{4}V_{6}).\]
The non-primitive charges for $l \ge 2$ are not commuting also.
\section{Conclusions}
In present paper author used dynamical polynomial chiral invariant charges to construct new
integrable equations. It is possible for $SU(3)$, $SO(3)$, $SP(2)$ groups only. The non-primitive
charges for groups of bigger rank are not Casimirs.

 This work was  supported by Grant of RFFI-NASU 38/50-2008.

\end{document}